# Functional and spatial rewiring principles jointly regulate context-sensitive computation

Jia Li[1*], Ilias Rentzeperis[1], Cees van Leeuwen[1,2]

[1]Brain and Cognition unit, Faculty of psychology and educational sciences, KU Leuven, Leuven, Belgium
[2]Cognitive and developmental psychology unit, Faculty of social science, University of Kaiserslautern, Kaiserslautern, Germany

*Corresponding author
Email: jia.li@kuleuven.be

**Abstract**

Adaptive rewiring is regarded as a general principle of self-organization in random neural networks. By selectively adding connections to regions with intense flow while pruning underused ones, adaptive rewiring regimes generate optimized, brain-like network topologies with modular, small-world, and rich club structures. Beyond topology, neural self-organization can also operate under certain spatial optimization principles, two prominent ones being the minimization of wiring distances and the maximization of aligned wiring layout. We simulated the interplay of rewiring based on these spatial principles (spatial rewiring) and adaptive rewiring in evolving neural networks with weighted and directed connections. The neural traffic flow within the network is represented by the equivalent of diffusion dynamics for directed edges: consensus and advection. We find that there is a constructive synergy between adaptive and spatial rewiring which contributes to network connectedness. More specifically, wiring cost minimization facilitates adaptive rewiring in creating convergent-divergent units. These structures have been hypothesized to aid in the flow of neural information and, in particular, to enable context-sensitive information processing in the sensory cortex. Each unit consists of convergent hub nodes which collect inputs from a pool of local nodes and project them via a densely interconnected set of intermediate nodes onto divergent hub nodes, which broadcast their output back to the pool of local nodes. The relative importance of spatial versus adaptive rewiring parametrically determines the degree to which the intermediate nodes are isolated from the rest of the network, and hence the context-sensitivity of its processing style.

246 words

## Introduction

Over the life cycle of the brain, neuronal connections are constantly changing (Butz, Wörgötter, van Ooyen, 2009; Chklovskii, Mel, Svoboda, 2004; Knott & Holtmaat, 2008). For nearly two decades, this ongoing evolution has been modeled by an elementary principle known as adaptive rewiring (Papadopoulos, Kim, Kurths et al., 2017, Gong & van Leeuwen, 2003; Fig 1.A). Adaptive rewiring facilitates signal processing by attaching shortcut connections to regions where neural signal traffic is intense while pruning underused connections.

Adaptive rewiring leads to complex topological features such as scale-freeness (Gong & van Leeuwen, 2003), small worlds (Gong & van Leeuwen, 2004), modularity (Rubinov et al., 2009; van den Berg & van Leeuwen, 2004), and the rich club effect (Hellrigel, Jarman, & van Leeuwen, 2019). These features reflect some of the most distinctive macroscopic characteristics of brain anatomy (for the small world, see Bullmore & Sporns, 2009; for the rich club effect, see Zamora-López, Zhou, & Kurths, 2010; van den Heuvel & Sporns, 2011).

Adaptive rewiring establishes these brain-like structure in models of oscillatory neural mass activity (Gong & van Leeuwen, 2003, 2004) as well as in neuronal-level spiking networks (Kwok, Jurica, Raffone, et al, 2007). Spike propagation in neural networks can be represented as random walks on a graph (Gong & van Leeuwen, 2009), and these, in turn, can be stochastically described as graph diffusion (Abdelnour, Voss, & Raj, 2014). Graph diffusion offers a particularly parsimonious account of neural activity, suitable for implementing adaptive rewiring in neural network models (Jarman, Steur, Trengove, et al. 2017).

Current adaptive rewiring models are still too simplified for representing the anatomical networks of the brain because they are exclusively concerned with optimizing topological features, while ignoring the spatial economy of the brain. For instance, a prominent spatial feature of the brain is that adjacent regions tend to have similar functions, and neural connections are aligned in fiber bundles and layers. This indicates that besides being adaptive, rewiring also follows certain spatial optimization principles.

One spatial principle involves minimization of wiring distance (Cherniak, 1994) by topologically connecting spatially adjacent nodes (Fig 1.B). Naturally, this principle pushes for the elimination of long-range connections, but, due to adaptive rewiring, a stable, albeit sparse, proportion of them remains (Jarman, Trengove, Steur, et al., 2014). These long-range connections preferentially attach to hub nodes, while short-distance connections are assigned to nodes within the same topological modules. The differentiation into long- and short-range connections evolves gradually, similarly to what happens in the developing brain (Oldham & Fornito, 2019). This results in a rewired network akin to a topographical map, a widespread functional architecture in the brain, found from the visual to the somatosensory cortex.

Alignment is another spatial principle (Fig 1.C) where connections tend to extend either in the same direction, as the axons of pyramidal cells in the cortex, or spread in a concentric fashion, as the dendrites of ganglia. A possible mechanism for alignment is that neuronal extensions develop along a vector field; either a chemical gradient or a traveling electrical wave field (Alexander, Jurica, Trengove et al., 2013; Muller, Chavane, Reynolds, & Sejnowski, 2018). Propagating wave fields have been proposed to play an active role in shaping cortical maps (Alexander, Trengove, Sheridan, et al. 2011). To the extent that wave fields are homogeneous or vary smoothly across spatial regions, rewired connections that align with the direction of the wave field tend to become spatially aligned with each other. Depending on the organization of the wave field, regular topography may arise, e.g., layers as a result of a homogeneously lateral wave or ganglia as a result of a radially expanding wave (Calvo Tapia et al., 2019).

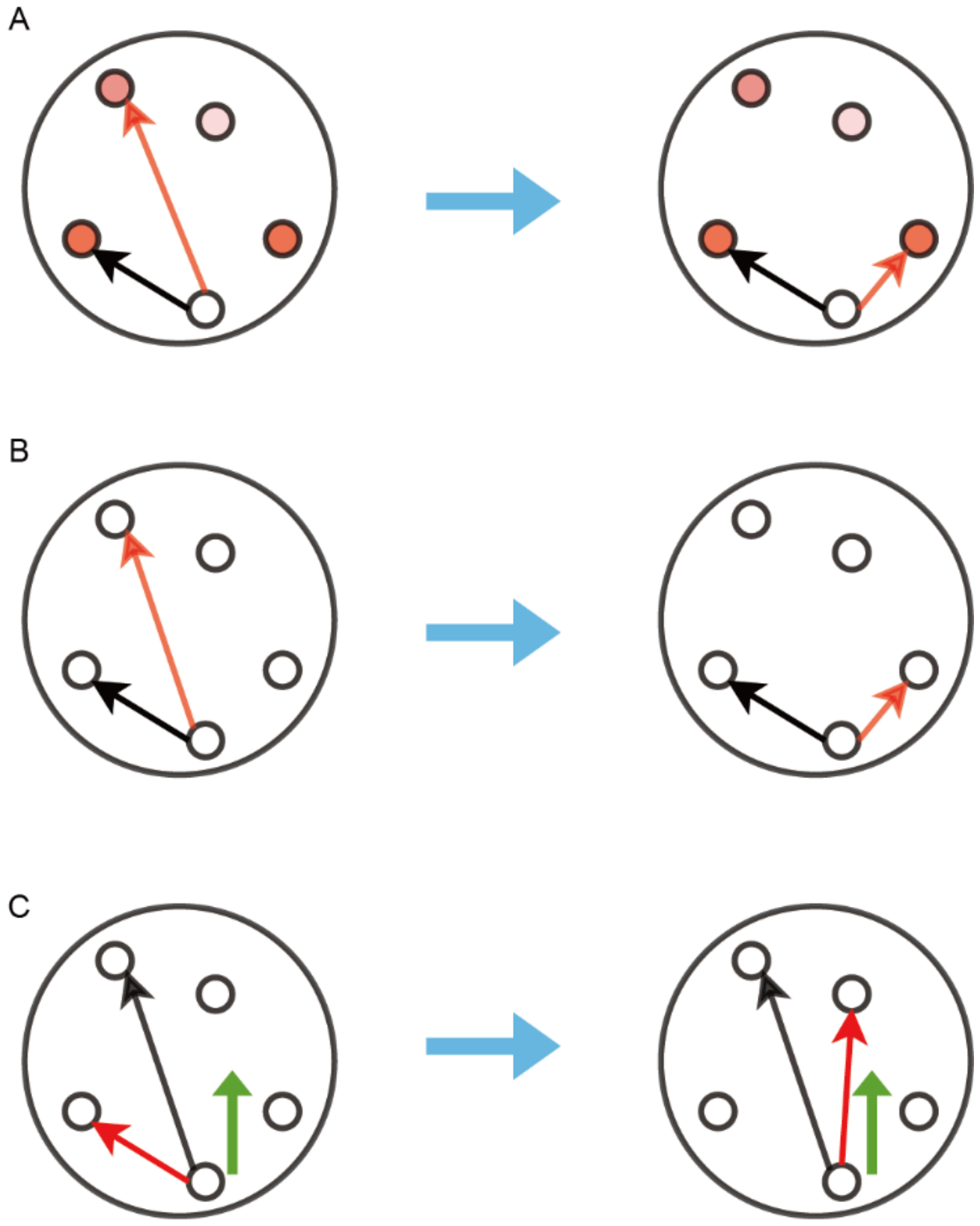

**Fig 1.** Principles of network rewiring. (A) Adaptive rewiring. The lightness of a node's color represents the intensity of its communication with the white node. The darker the color, the more intense the communication is (B) Minimization of wiring distance. (C) Alignment to an external vector field. The red and green arrows indicate the rewiring link and the direction of the vector field respectively.

In a recent study on undirected binary networks that incorporated both spatial principles, the emerging networks revealed morphologies that are stalwarts of the nervous system's functional anatomy, such as parallelism, super-rings, and super-chains, while they maintained the complex network properties generated by adaptive rewiring (Calvo Tapia et al., 2019).

Recent model developments have enabled extensions of the adaptive rewiring principle to undirected weighted (Rentzeperis & van Leeuwen, 2020; Rentzeperis & van Leeuwen, 2021) and directed binary graphs (Rentzeperis, Laquitaine, & van Leeuwen, 2022). Here we use directed, weighted networks while incorporating both spatial principles.

Complementary to Calvo Tapia et al. (2019), who were concerned with network morphology, we focus on the evolution in these networks of a particular kind of structure that facilitates context-sensitive computation. In biological networks, context-sensitive computation is efficiently achieved through pooling, i.e., certain hub units collect converging inputs, and pass this information to divergent output-hubs via subnetworks of intermediate nodes (Fig 2). Such structures are known as convergent-divergent units (Kumar, Rotter, & Aertsen, 2010; Shaw, Harth, & Scheibel, 1982). Prominent examples of convergent-divergent units are the circuits in V1 underlying contextual modulation: pools of orientation selective neurons in layers 2/3 that send their input to somatostatin (SOM) cells, which then broadcast back to the pool of neurons their response (Adesnik et al., 2012; Niell & Scanziani, 2021). The SOM hub cells form with the vasoactive intestinal peptide (VIP) neurons an intermediate unit between convergence and divergence that adjusts the contextual modulation response of the pool of neurons as the relationship between surround and stimulus changes (Keller et al., 2020). At a different scale and for a different scope, the cortico-basal ganglia circuitry can be seen as a convergent-divergent units that regulates voluntary movement: the striatum receives multimodal contextual information from the cortex, processes it and sends it to other subcortical structures such as the pallidum and the substantia nigra, and then the thalamus which acts as a divergent hub broadcasts the processed output back to the cortex (Redgrave et al., 2010).

Convergent-divergent units thus constitute the connective core of sensory, motor, and cognitive brain regions (see Krause & Pack, 2014 for a review), and of global networks (Shanahan, 2012). They allow the receptive fields of sensory neurons to still be driven by local features but modulated by global contextual features (Keller, Dipoppa, Roth, et al., 2020). This enables, among other things, surround suppression via connections within area V1 (Das, & Gilbert, 1999; Hupé, James, Payne, et al., 1998) and sensorimotor prediction coding via long-range connections onto the visual system (Jordan, & Keller, 2020, Leinweber, Ward, Sobczak, et al., 2017; Keller, Bonhoeffer, & Hübener, 2012).

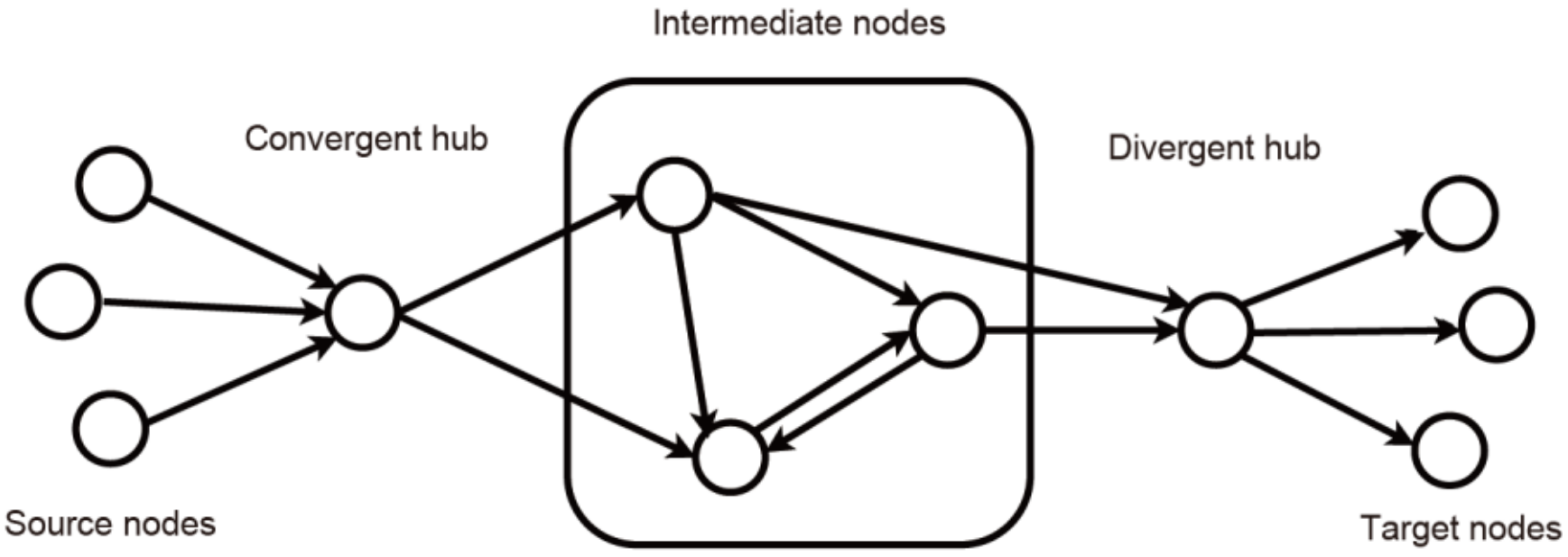

**Fig 2.** Schema of a convergent-divergent unit. In a convergent-divergent unit, a convergent hub collects inputs and passes the information to a divergent hub through a subnetwork of intermediate nodes. The nodes sending information to the convergent hub are referred as source nodes, and those receiving information from the divergent hub as target nodes. Note that typically the source and target nodes are overlapping, i.e., a node can be both a source and a target node.

By employing adaptive rewiring to (binary) directed graphs, Rentzeperis et al., (2022) observed in their model the emergence of convergent-divergent units. Here we use directed weighted graphs to study the effect of spatial optimization principles, i.e., distance minimization and alignment, on the development of convergent-divergent units. We find that the distance minimization principle enables nodes to be encapsulated within convergent-divergent units. The alignment principle interferes with the formation of the convergent-divergent units, the extent of which depends on the type of the layout.

Context-sensitivity differs across brain regions: more local for early and mid-level visual areas (Field, Hayes, & Hess 1993) and more global for higher order ones (Quian Quiroga, Reddy, Kreiman, et al.,2005). It also may vary between individuals (Yamashita, Fujimura, Katahira, et al., 2016), the sexes (Phillips, Chapman, & Berry, 2004), and cultural groups (Doherty, Tsuji, & Phillips, 2008). These style variations might be associated with variations in the convergent-divergent units. In our models, we find that the degree to which nodes within convergent-divergent units are encapsulated, or isolated from the rest of the network, depends parametrically on the relative prominence of distance minimization and adaptive rewiring.

## Methods

### Notation and definitions

A directed graph (digraph) is described by the set, $G = (V, E, W)$, where $V = \{1, 2, \ldots, n\}$ is the set of nodes, $E \subset V \times V$ the set of ordered node pairs with $(i, j) \in E$ representing directed edges from $i$ to $j$ denoted as $i \to j$, and $W = \{w_{ij}: i, j \in V\}$ the set of edge weights, where $w_{ij} > 0$ if $(j, i) \in E$, and $w_{ij} = 0$ when $(j, i) \notin E$. For a weighted edge $j \to i$, its length equals its inverse weight $\frac{1}{w_{ij}}$. The cardinalities $|V| = n$ and $|E| = m$ denote the numbers of nodes and directed edges respectively.

Nodes are called adjacent if there is an edge (in either direction) between them. The $n \times n$ adjacency matrix $A = \left[A_{ij}\right]_{i,j \in V}$ carries the edge weights of a network as $A_{ij} = w_{ij}$ (Fig 3). We refer to the edges directed at node $i \in V$ as the in-link of $i$ and the edges starting from node $i$ as the out-link of $i$. The tails of the in-links of $i$ constitute the in-degree neighborhood of $i$, $N_{in}(i)$. The remaining set of nodes, $V - i - N_{in}(i)$, is denoted as $N_{in}^{c}(i)$. The in-degree of node $i$ is the number of its in-links. Analogously, the heads of the out-links of $i$ constitute the out-degree neighborhood of $i$, $N_{out}(i)$, and the rest is denoted as $N_{out}^{c}(i)$. The out-degree of node $i$ is the number of its out-links. For an ordered node pair $(u, v)$, a directed walk from $u$ to $v$ is an ordered list of edges $\{(i_0, i_1), (i_1, i_2), \dots, (i_{K-1}, i_K): i_0 = u, i_K = v, (i_{k-1}, i_k) \in E\}$ (Bender & Williamson, 2010). A directed walk is a directed path if the vertices on it are distinct.

**Fig 3.** Schema of the adjacency matrix. The elements of the adjacency matrix are the weights of links. Each row of the adjacency matrix contains the weights of in-links for the corresponding node, and the number of nonzero entries is the in-degree. Similarly, each column carries the weights of out-links, and the number of nonzero entries is the out-degree.

### Consensus and advection dynamics

Rentzeperis et al. (2022) generalized the diffusion dynamics used for undirected graphs in Jarman et al. (2017) to consensus and advection dynamics in digraphs. Both consensus (Ren et al., 2007) and advection dynamics (Chapman, 2015) drive nodes' values to converge to a global state based on each node's local state. The local state of each unit is described by a value, which is called its concentration. Communication between nodes is represented as in- and out-flow between adjacent nodes. Consensus and advection algorithms assume different dynamics driving the communication between nodes.

In consensus dynamics there is communication between adjacent nodes if they have different concentrations. The rate of change of the concentration of node $i$ under the consensus dynamics is

$$\dot{x}_i(t) = \sum_{\{\forall j | j \to i\}} w_{ij} \left( x_j(t) - x_i(t) \right) \tag{1}$$

The in-degree Laplacian matrix, $L_{in} = \{l_{ij}^{in}\}$, of the graph $G$ is defined as

$$l_{ij}^{in} = \begin{cases} \sum_{k=1}^{n} w_{ik}, if\ i = j \\ -w_{ij}, if\ i \neq j \end{cases} \tag{2}$$

We substitute (2) into (1), and the consensus dynamics in matrix form becomes:

$$\dot{x}(t) = -L_{in}x(t) \tag{3}$$

The nodes' concentration at time t is then:

$$x(t) = e^{-L_{in}t}x(0) \tag{4}$$

In the case of advection dynamics, a node receives an inflow from its in-degree neighbors and sends an outflow to its out-degree neighbors. The inflow and outflow are analogous to the weight of the links and the concentrations of the nodes. Mathematically, the advection dynamics for node $i$ is

$$\dot{x}_i(t) = \sum_{\{\forall j|j\to i\}} w_{ij}x_j(t) - \sum_{\{\forall k|i\to k\}} w_{ki}x_i(t) \tag{5}$$

The out-degree Laplacian matrix, $L_{out} = \{l_{ij}^{out}\}$, of the graph $G$ can be defined as

$$l_{ij}^{out} = \begin{cases} \sum_{k=1}^{n} w_{ki}, if\ i = j \\ -w_{ij}, if\ i \neq j \end{cases} \tag{6}$$

The advection dynamics in matrix form is

$$\dot{x}(t) = -L_{out}x(t) \tag{7}$$

and the solution is

$$x(t) = e^{-L_{out}t}x(0) \tag{8}$$

For each node, we set its initial concentration to be one and the rest to be zero. The initial conditions collectively form an identity matrix. Applying these initial conditions to eq (4) and eq (8), we get

$$c(t) = e^{-L_{in}t}I_{n\times n} = e^{-L_{in}t} \tag{9}$$

and

$$\alpha(t) = e^{-L_{out}t}I_{n\times n} = e^{-L_{out}t} \tag{10}$$

We refer to $c(t)$ and $a(t)$ as the consensus and advection kernels respectively. Both reflect the intensity of communication between nodes. For the purpose of our study, the time variable $t$ for the two kernels could also be thought of as a rewiring interval, indicating the time elapsed between two successive rewiring iterations. In all our experiments it was set to 1.

**Rewiring Principles**

We probe how the structure of the network changes when we iteratively rewire its edges. In general, at each iteration, a node $v \in V$ is randomly selected and depending on the rewiring condition either one of its in-link is cut and another added, or an out-link is cut and another added. Suppose that we are going to rewire an in-link of $v$ in an iteration step. An edge $(k, v) \in E$ will be dropped, and a new edge from $v$ to a node $l$ will be established where $v$ was not previously connected, $(l, v) \notin E$, will be added. When an out-link of $v$ is rewired, an edge $(v, k) \in E$ is substituted by a new edge $(v, l) \in E$.

To decide the choice of $k$ and $l$ at each rewiring step, one of the following three principles, explained below, will be selected with a fixed probability: either the functional principle of adaptive rewiring or one of the two spatial principles: the distance or the wave principle.

**Functional principle: Adaptive rewiring**

The adaptive rewiring principle is called a functional principle, as it depends on the activation flow between nodes. It states that an underused connection is removed and a new connection is established between two previously unconnected nodes with the most intense traffic between them (via all indirect paths). Distinct topological patterns develop when rewiring the in-degree neighborhood with the consensus algorithm and when rewiring the out-degree neighborhood with the advection algorithm (Rentzeperis et al., 2022). Therefore, we model the intensity of communication with the consensus kernel when rewiring the in-links and with advection kernel when rewiring the out-links of candidate nodes.

When an in-link of $v$ is rewired, $k$ is the node in $N_{in}(v)$ such that $(k, v)$ has the lowest consensus kernel value, i.e., $k = argmin_{u \in N_{in}(v)}\{c(t)_{vu}\}$ (the $(k, v)$ connection is cut), and $l$ is the node in $N_{in}^{c}(v)$ such that $(l, v)$ has the largest consensus kernel value, i.e., $l =$ $argmax_{u \in N_{in}^c(v)}\{c(t)_{vu}\}$. In a similar fashion, when rewiring an out-link, $k$ is the node in $N_{out}(v)$ such that $(v, k)$ has the lowest advection kernel value, i.e., $k = argmin_{u \in N_{out}(v)}\{a(t)_{uv}\}$, and $l$ is the node in $N_{out}^c(v)$ such that $(v, l)$ has the largest advection kernel value, i.e., $l = argmax_{u \in N_{out}^c(v)}\{a(t)_{uv}\}$.

**Spatial principles: Spatial Rewiring**

To instantiate the remaining two principles, the digraph is embedded in a two-dimensional Euclidean space, where the coordinates of node $i$ is denoted as $\vec{x}_i \in R^2$.

**Distance principle.** According to the distance principle, the longest connection is removed and replaced by the spatially closest connection possible between two previously unconnected nodes. The distance between node $i$ and node $j$ is given by $d_{ij} = ||\vec{x}_i - \vec{x}_j||$, where $|| \cdot ||$ is the Euclidean distance. When an in-link of $v$ is rewired, $k$ is the node in $N_{in}(v)$ with the longest distance from $v$, i.e., $k = argmax_{u \in N_{in}(v)}\{d_{uv}\}$, and $l$ the node in $N_{in}^c(v)$ with the shortest distance from $v$, i.e., $l = argmin_{u \in N_{in}^c(v)}\{d_{uv}\}$. When rewiring an out-link of $v$, $k$ is $argmax_{u \in N_{out}(v)}\{d_{vu}\}$ and $l$ is $argmin_{u \in N_{out}^c(v)}\{d_{vu}\}$.

**Wave principle.** The wave principle serves to optimize spatial alignment between network connections. It removes the connection at the largest angle to a vector field $\vec{F}(\vec{x})$ and replaces it with a connection between two previously unconnected nodes whose direction is most closely aligned with the direction of the vector field. The cosine of the angle between the edge $j \rightarrow i$ and the vector field at $\vec{x}_v$ is $\cos(\theta_{ij}) = \frac{(\vec{x}_i - \vec{x}_j) \cdot \vec{F}(\vec{x}_v)}{d_{ij}||\vec{F}(\vec{x}_v)||}$, $\theta_{ij} \in [0, \pi]$. When rewiring an in-link of $v$, $k$ is the node in $N_{in}(v)$ such that $(k, v)$ forms the largest angle with $\vec{F}(\vec{x}_v)$, i.e., $k = argmin_{u \in N_{in}(v)}\{cos(\theta_{uv})\}$, and $l$ is the node in $N_{in}^c(v)$ such that $(l, v)$ forms the smallest angle with $\vec{F}(\vec{x}_v)$, i.e., $l = argmax_{u \in N_{in}^c(v)}\{cos(\theta_{uv})\}$. When rewiring an out-link of $v$, $k$ is $argmin_{u \in N_{out}(v)}\{\cos(\theta_{vu})\}$ and $l$ is $argmax_{u \in N_{out}^c(v)}\{cos(\theta_{vu})\}$.

**Rewiring algorithm**

Throughout the rewiring process, the number of nodes and edges of the networks are kept constant for the sake of simplicity (but see Gong & van Leeuwen, 2003 for growing and van den Berg, Gong, Breakspear, et al., 2012 for pruning in undirected networks). The rewiring process starts from a random directed network $D = (V, E, W)$ with predetermined node number $n$ and edge number $m$. Edges are assigned to $m$ nodes pairs that are randomly selected from all $n(n-1)$ node pairs without replacement. Then positive weights sampled from a predetermined probability distribution are randomly assigned to these edges.

The iterative rewiring process proceeds as follows.

Step 1: Select a random node $v \in V$ such that its in- and out-degrees are neither zero nor $n-1$.

Step 2: With probability $p_{in}$ rewire an in-link of $v$ using consensus; otherwise, rewire an out-link of the same node using advection. When $p_{in}$ is set to 1, only in-links are rewired. In this case, we refer to the iterative process as in-link rewiring. Analogously, when $p_{in}$ is set to 0, we refer to the iterative process as out-link rewiring.

Step 3: Depending on the result in step 2, select a random in-link or out-link of $v$ and rewire it according to one of the three rewiring principles. The probabilities of choosing the distance principle, wave principle, or functional principle are $p_{distance}$, $p_{wave}$, and $p_{function}$ ($p_{function} = 1 - p_{distance} - p_{wave}$), respectively.

Step 4: Return to step 1 until $M$ edges have been rewired.

We refer to this algorithm as the 'functional + spatial' algorithm. To allow the reader a feel for the effects of functional and spatial rewiring principles, we provided illustrative examples in Supplementary Materials S1.

**Baseline algorithm**

We take the algorithm from Rentzeperis et al. (2022) as the baseline algorithm, in which the functional principle was combined with random rewiring. Random rewiring drops and adds links randomly. Suppose that an in-link of a node $v \in V$ is rewired. According to random rewiring, two nodes $k \in N_{in}(v)$ and $l \in N_{in}^{c}(v)$ are selected randomly. The in-link $(k, v)$ is cut and $(l, v)$ is added. Random rewiring for out-links is similar. The probabilities of choosing functional principle and random rewiring at each iteration are $p_{function}$ and $p_{random}$ ($p_{random} = 1 - p_{function}$), respectively.

This algorithm was originally applied to directed binary networks. We run this algorithm on directed weighted networks to test if the similar results are obtained and compare the effects of random rewiring to that of spatial principles. This algorithm is referred as the 'functional + random' algorithm.

**Network measures**

To study the impact of the rewiring principles on the structure of weighted digraphs, we calculate following measures for each rewired network. High scores on each of these measures reflect better information processing and communication within the network.

**Number of connected node pairs.** An ordered pair $(i, j)$ is connected if there is a directed path from $j$ to $i$. The number of connected node pairs is a measure of the extent of information exchange in a digraph. The upper bound of the number of connected node pairs is $n^2$ which is achieved when every node can send information to any node, including itself. We use this measure to quantify the connectedness of a digraph.

**Average efficiency.** The average efficiency measure quantifies the efficiency of sending information over a network, which is defined as the mean of the inversed shortest directed path lengths of all node pairs (Latora & Marchiori, 2001):

$$E = \frac{1}{n(n-1)} \sum_{i \neq j \in V} \frac{1}{d_{ij}} \quad (11)$$

where $d_{ij}$ is the length of the shortest directed path from node $i$ to node $j$. If there is no directed path from $i$ to $j$, $d_{ij} = \infty$.

**Number of hubs.** We define convergent hubs as nodes with at least one out-link and a number of in-links that are above a threshold we define. These hubs are suitable as a substrate for collecting distributed information. Inversely, divergent hubs are nodes with at least one in-link and a number of out-links above a predefined threshold. These hubs are suitable as a substrate for information broadcasting. The threshold was set to 15 for both convergent and divergent hubs in the following analysis.

**Simulation parameter settings**

In our simulations, the number of nodes was $n$ = 100. We set the number of edges to $m = [2(n) * (n-1)] = 912$, a number sufficiently low for a network to be considered sparse (cf. van den Berg, et al, 2012 for undirected networks). The unnormalized weights were either sampled from a normal, $N(1, 0.25^2)$ or a lognormal distribution $logN(0,1)$. Sampled negative weights from the normal distribution (an almost impossible occurrence as indicated by its probability: $3.17 * 10^{-5}$) were set to 0.05. Normalized weights were obtained by dividing the sampled weights by the sum of all weights. That way the sum of the new normalized weights equals to the number of edges.

For the purposes of spatial embedding, nodes were placed randomly with a uniform distribution on a unit disk; the external field was set to a field $\vec{F}(\vec{x}) = (1,0)$ to induce parallel connections or a radial field $\vec{F}(\vec{x}) = \frac{\vec{x}}{\|\vec{x}\|}$ to induce concentric connections.

The number of iterations $M$ for each run was 4000. The probabilities of the three principles $(p_{function}, p_{distance}, p_{wave})$ and the probability of rewiring in-link $p_{in}$ were kept fixed for each run but varied between simulations. For each combination of parameters $(p_{in}, p_{function}, p_{distance}, p_{wave})$, we run 30 different instantiations of the rewiring algorithm, over which the mean and standard deviation of the measures were calculated.

## Results

We first test whether a combination of adaptive and random rewiring can produce a convergent-divergent unit when the weights of the connections follow a normal distribution. Then we test whether the distance principle facilitates in the formation of convergent-divergent units similarly to random rewiring. Finally, we probe the effect of the wave on the formation of convergent-divergent units. We found similar results when the weights of the connections followed a lognormal distribution, which are not shown here.

**Convergent-divergent units on directed weighted networks**

We run the 'functional + random' algorithm on directed weighted networks for various $(p_{function}, p_{random})$.

As the name implies, the emergence of convergent-divergent units requires the formation of both convergent and divergent hubs as well as the existence of communication pathways between them. We find that the number of connected node pairs increases with the proportion of random rewiring, $p_{random}$, regardless of the proportion of in-link rewiring, $p_{in}$ (Fig 4.A). The effect of $p_{random}$ on efficiency (Fig S2.A) is consistent with that on connectedness, which implies the increase of the average efficiency results from the increase of the connectedness of the networks.

As expected, the number of convergent hubs decreases and the number of divergent hubs increases when $p_{in}$ increases (Fig 4.B, C). When both in-link and out-link are rewired, i.e., $0 < p_{in} < 1$, the number of convergent and divergent hubs peaks at intermediate values of $p_{random}$ (Fig 4.B, C).

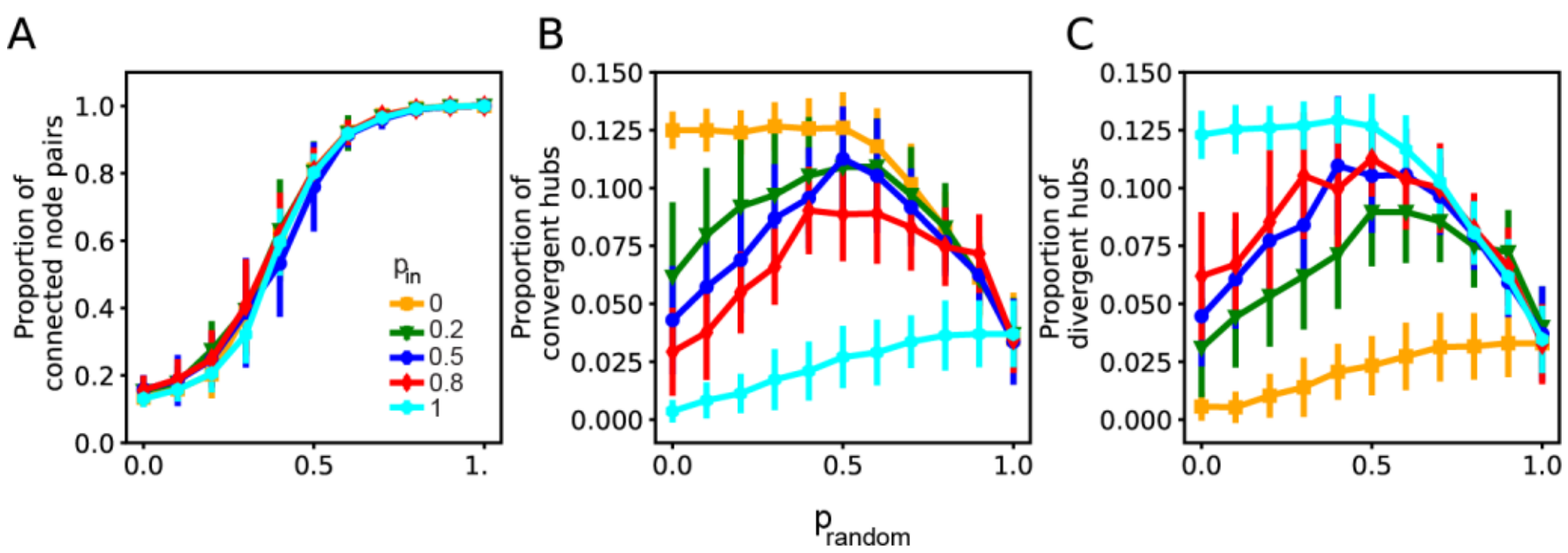

**Fig 4.** Random rewiring enhances connectedness and increases the number of hubs when rewiring includes both advection and consensus ($0 < p_{in} < 1$). (A) The proportion of connected node pairs as a function of the probability of random rewiring, $p_{random}$, for different in-link rewiring probabilities, $p_{in}$. (B) The proportion of convergent hubs as a function of $p_{random}$. (C) The proportion of divergent hubs as a function of $p_{random}$.

We examine whether the convergent and divergent hubs in the network are connected in such a way so that they form convergent-divergent units. The probability of rewiring in-links, $p_{in}$, is set to 0.5, so that equal proportions of convergent and divergent hubs could develop. Convergent-divergent units emerge in directed networks when adaptive and random rewiring are combined (Fig 5.A, B). The number of convergent-divergent units initially increases with $p_{random}$, but then drops for $p_{random} > 0.6$ (Fig 5.B). Although convergent and divergent hubs exist when $p_{random} = 0$ (Fig 4.B, C), they hardly connect to each other, making the number of convergent-divergent units close to 0 (Fig 5.B).

The connectedness of the hub nodes also depends on the proportion of random rewiring. For a convergent-divergent unit, we refer to nodes that can be reached from the divergent hub as the target nodes and the nodes that send information to the convergent hub as the source nodes (Fig 2). The source nodes and target nodes are allowed to overlap. The proportions of source and target nodes to the whole network increase with the probability of random rewiring, $p_{random}$ (Fig 5.C). The proportion of their overlap also increases with $p_{random}$ (Fig 5.C).

For each convergent-divergent unit, we refer to nodes on directed paths from the convergent to the divergent node as intermediate nodes, and to the subgraph which consists of these nodes as the intermediate subgraph. The intermediate subgraph processes the information collected by the convergent hub. The degree of its isolation from the rest of the network characterizes the context-sensitivity of its processing style. We calculated for all convergent-divergent node units, the size and density of the intermediate subgraph, as long as this subgraph contained more than one node. For each combination of $(p_{function}, p_{random})$, the sizes and densities of the subgraphs were pooled together across 30 instances.

We found that the size of the subgraph increases with $p_{random}$ (Fig 5.D) while the average density decreases until it reaches and stays at a floor at $p_{random} > 0.5$, near the density of the whole digraph (Fig 5.E). This is because the intermediate subgraph contains all of the nodes of the graph except for the convergent and divergent hubs.

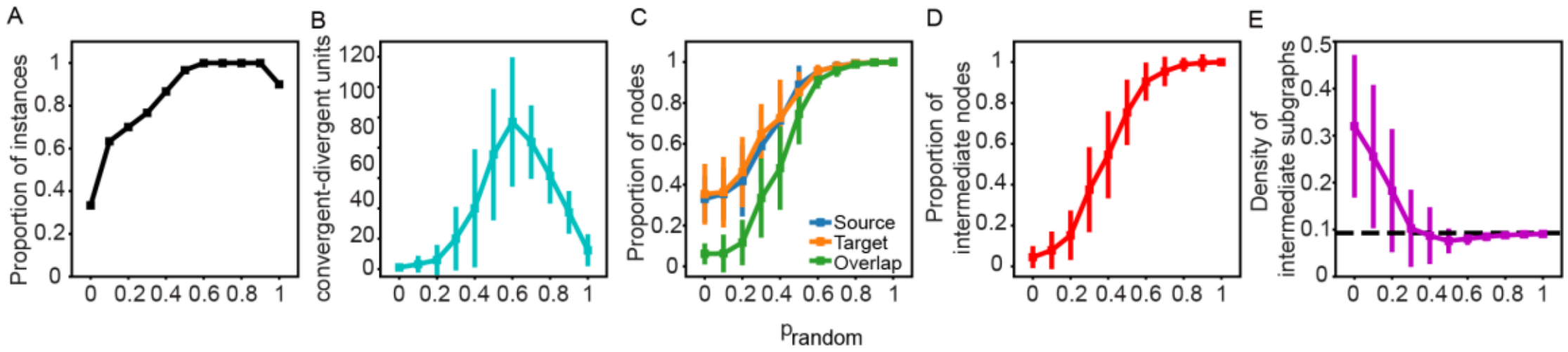

**Fig 5.** $p_{random}$ controls the formation, connectedness, and the degree of isolation of convergent-divergent units. (A) Proportion of rewired networks forming convergent-divergent units as a function of $p_{random}$. (B) Number of convergent-divergent units in rewired networks as a function of $p_{random}$. (C) Proportion of source nodes, target nodes and their overlap as a function of $p_{random}$. (D) Proportion of nodes in intermediate subgraphs as a function of the proportion of $p_{random}$. (F) Density of intermediate subgraphs as a function of $p_{random}$. The black horizontal line is the density of the whole digraph.

**Similar effects of distance-based rewiring as random rewiring**

To see the effects of distance-based rewiring, we replace random rewiring with distance-based rewiring. We run the 'functional + spatial' algorithm without including wave-based rewiring ($p_{wave} = 0$). Note that the 'functional + spatial' algorithm does not include any random rewiring.

We found that the proportion of distance-based rewiring, $p_{distance}$, has a similar effect on connectedness, average efficiency and hub formation in the rewired networks as random rewiring (compare Fig 6 with Fig 3, and Fig S2.B with Fig S2.A).

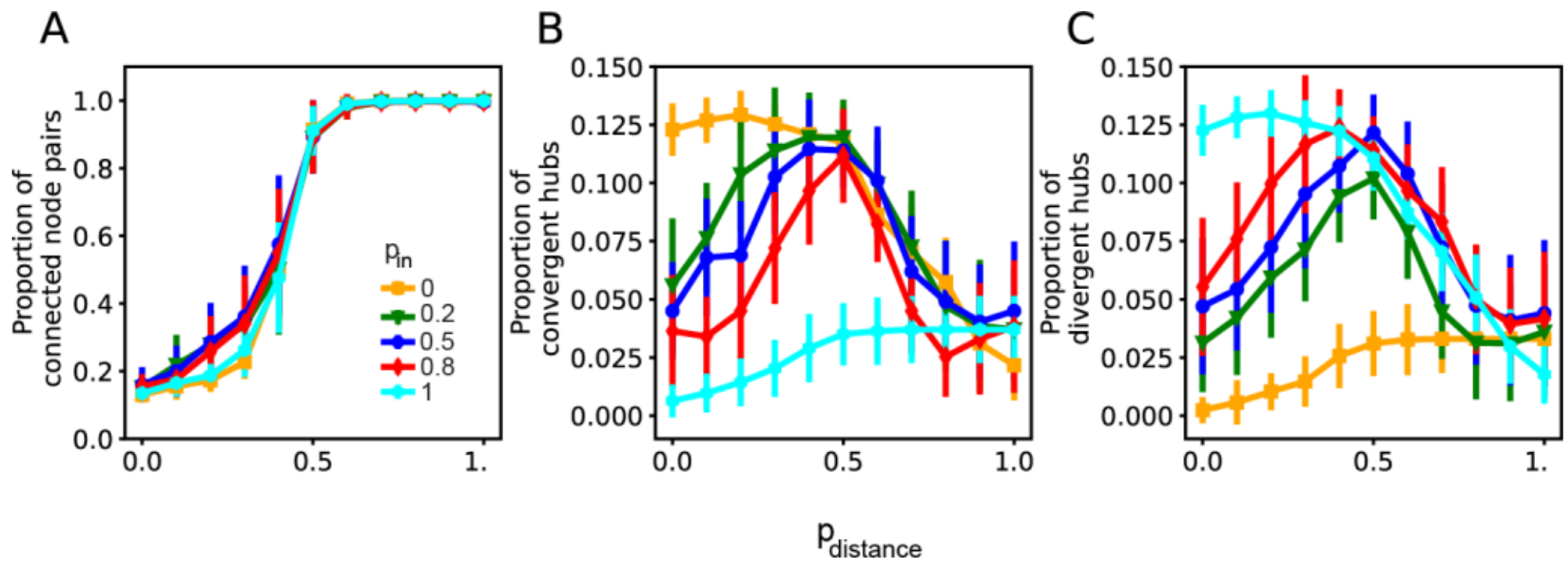

**Fig 6.** Distance-based rewiring has similar effects on the connectedness and the number of hubs as random rewiring. (A) The proportion of connected node pairs as a function of the probability of distance-based rewiring, $p_{distance}$, for different probabilities of in-link rewiring, $p_{in}$. (B) The proportion of convergent hubs as a function of $p_{distance}$. (C) The proportion of divergent hubs as a function of $p_{distance}$.

The success rate of the formation of convergent-divergent units slightly decreases compared to using random rewiring (Fig 7.A), while other statistics are similar to those in random rewiring (Fig 7.B-E). The distance principle affects the style of convergent-divergent units similarly to random rewiring. The average density decreases until $p_{distance} \geq 0.5$ , then remains stable (Fig 7.E). We may conclude that, as long as $p_{distance} < 0.5$ , we can regard the intermediate nodes as encapsulated to various degrees from the rest of the network, regulated by $p_{distance}$. This effect is independent of the modularity of the network structure, which changes proportionally with $p_{distance}$ (Fig S3). The present result indicates that random rewiring can effectively be replaced by spatially-based rewiring according to the distance principle.

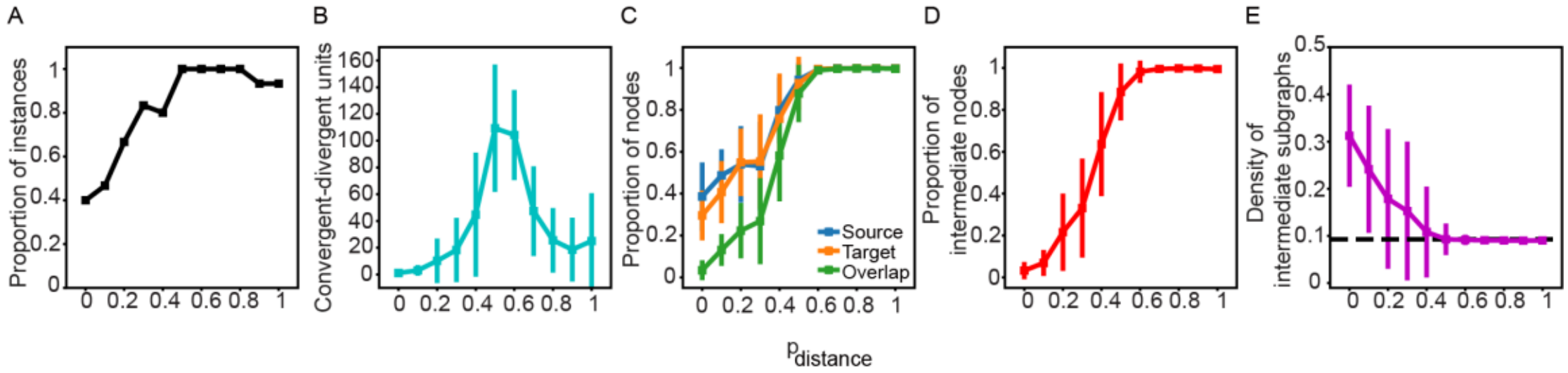

**Fig 7.** $p_{distance}$, controls the formation, connectedness, and the degree of isolation of convergent-divergent units. (A) Proportion of rewired networks forming convergent-divergent units as a function of $p_{distance}$. (B) Number of convergent-divergent units in rewired networks as a function of $p_{distance}$. (C) Proportion of source nodes, target nodes and their overlap as a function of $p_{distance}$. (D) Proportion of nodes in intermediate subgraphs as a function of the proportion of $p_{distance}$. (E) Density of intermediate subgraphs as a function of $p_{distance}$. The black horizontal line represents the density of the whole digraph.

**Effects of wave principle on the formation convergent-divergent units**

When the wave principle was included, the networks developed different layouts corresponding to the underlying field (see the logos in Fig 8). The wave-based rewiring does not affect the effect of distance-based rewiring on the connectedness and hub formation (Fig S4.1).

The formation of convergent-divergent units is affected by the wave principle (Fig 8). For the lateral field case, the proportion of rewired networks that contains convergent-divergent units could become small at some combinations of $(p_{function}, p_{distance}, p_{wave})$ when $p_{distance} < 0.5$ (Fig 8.A); while for the radial field case, this number is high across all combinations of $(p_{function}, p_{distance}, p_{wave})$ (Fig 8.B). The other statistics of the convergent-divergent units change with the proportion of distance-based rewiring in a similar way as without wave-based rewiring (Fig S4.2-4).

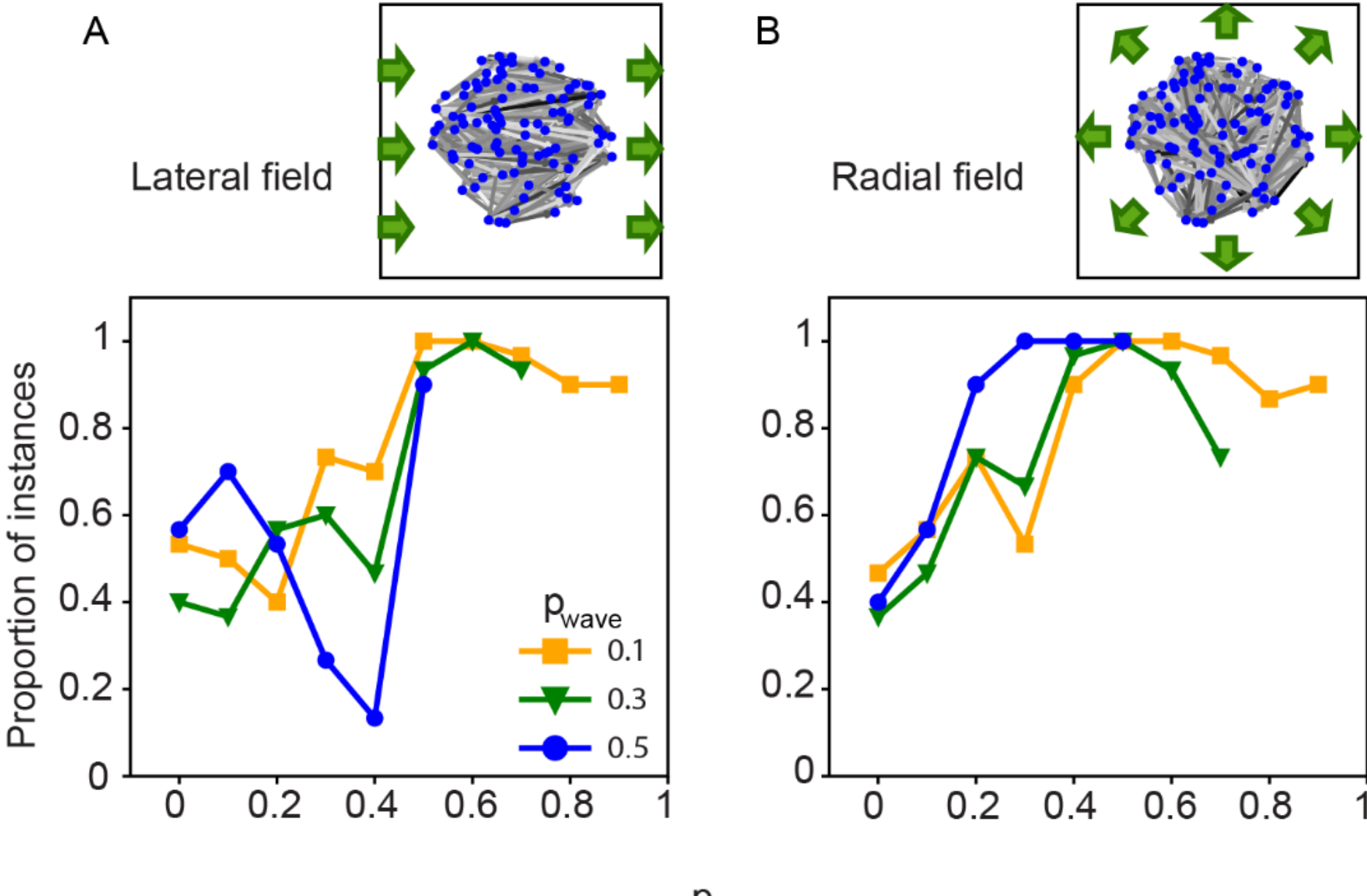

**Fig 8.** The wave principle affects the formation of the convergent-divergent units. Proportion of rewired networks forming convergent-divergent units as a function of the proportion of distance-based principle, $p_{distance}$, (A) for the lateral field case and (B) radial field case. The logos next to titles are the morphological view of the networks when the wave propagates laterally and radially. The proportion of in-link rewiring is 0.5, and $(p_{function}, p_{distance}, p_{wave})$ is $(0.4, 0.3, 0.3)$. The green arrows indicate the direction of the underlying field.

## Discussion

Starting from a random graph, repeated adaptive rewiring leads to complex network structures. Previous studies explored this phenomenon for directed binary and undirected graphs; here we extended the scope of this principle to directed weighted graphs and considered their spatial embedding (Calvo Tapia et al., 2019). Similar to these studies, at each rewiring step, we randomly chose with different proportions from three basic rewiring principles: the functional principle of adaptive rewiring according to its ongoing network activity, as represented here by generalized diffusion (i.e., advection and consensus dynamics; Rentzeperis et al., 2022) and two spatial principles: wiring distance minimization and vector field alignment. All three principles are deemed important in shaping the morphology of the nervous system (Calvo Tapia et al, 2019).

We found that functional and spatial principles took complementary roles: whereas adaptive rewiring principle takes the role of forming hubs, the distance minimization principle ruled over network connectedness and efficiency. Previous studies in adaptive rewiring without spatial principles found that adaptive rewiring only, while effective in forming hubs and modules, tends to reduce connectivity and efficiency (Rentzeperis et al., 2022). In their seminal study, Watts &

Strogatz (1998) showed that adding a small proportion of random connections strongly improved the efficiency and connectedness of a modular network. For this reason, all previous adaptive rewiring studies, from Gong & van Leeuwen (2003) to Rentzeperis et al. (2022), introduced a proportion of random rewiring to their network models, thus securing efficiency and connectedness. As spatial rewiring principles are shown to play a similar role, they successfully substitute random rewiring in our model. A major difference between random and spatial rewiring is that the former benefits global connectivity (Watts & Strogatz, 1998), whereas the latter favors local connectivity (Cherniak, 1994). This discrepancy, however, was shown to be no obstacle to the formation of hubs in the network. In fact, applying the distance principle showed the network to evolve a modular structure.

A limitation of the current study is that in our models, the relative contribution of all three rewiring principles was fixed during the network evolution. We did not consider the possibility that different rewiring principles change in prominence over time. Early in the development of the brain, vector field alignment may play a rather prominent role, as brain activity around gestation shows massive bursts of action potentials that spread in a wave-like manner (Blankenship & Feller, 2010). By contrast, the formation of hubs continues over a much longer period that extends into late adolescence (Oldham & Fornito, 2019).

In directed networks, we may distinguish convergent hubs and divergent hubs. Which of these is more prominent depends on the proportion of advection and consensus dynamics applied in the model (Rentzeperis et al., 2022). This feature may be useful to customize networks to processing requirements, e.g., divergence may be more useful in early processing regions; convergence in later ones (Gorban, Makarov, & Tyukin, 2019). When the advection and consensus dynamics are balanced, adaptive rewiring forms equal numbers of convergent and divergent hubs. These are the major constituents of convergent-divergent units. Convergent-divergent units collect information from pools of nodes through the convergent hubs, process the information in intermediate nodes, and broadcast the results to the network through the divergent hubs. In the brain these units enable context-sensitive modulation of network activity. In the model, convergent-divergent units are formed when convergent and divergent hubs arise (due to adaptive rewiring) and, when the network is efficiently connected (due to the distance minimization principle). We found that, as long as adaptive rewiring and the distance minimization principle are balanced in the evolving network, convergent-divergent hubs are successfully formed.

The distance minimization principle thus interacted constructively with adaptive rewiring in the formation of convergent-divergent units. Moreover, it contributed modularity to the network and established a rich club effect amongst the hubs. Because of this, the units jointly constitute the connective core (Shanahan, 2012) of the network.

An important feature of the distance minimization principle is that its prominence in rewiring determines the degree of encapsulation of the intermediate nodes in the convergent units. With lower proportions of distance-based rewiring, the intermediate nodes were relatively isolated from the rest of the network; with higher proportions they were more interconnected with it. In other words, the relative contribution of the distance principle regulates the context-sensitivity of the computations performed in the convergent-divergent units. We may consider the possibility that this feature is used to establish that whole networks differ regarding the context-sensitivity of their processing style (Doherty, et al., 2008; Phillips et al., 2004; Yamashita, et al., 2016) or to tailor the convergent-divergent units of different subnetworks to their specific computational requirements (Field et al., 1993; Krause & Pack, 2014; Quian Quiroga et al.,2005).

Different types of convergent-divergent units could be assigned to different subnetworks. Several different subnetworks have been distinguished in the brain, such as the dorsal attention, (Corbetta, Kincade, Ollinger, et al., 2000; Hopfinger, Buonocore, & Mangun, 2000), the salience (Menon & Uddin, 2010; Seeley, Menon, Schatzberg, et al., 2007) and the default mode network (Raichle, 2015; Raichle, MacLeod, Snyder, et al., 2001) that could operate, in competition or in cooperation. The set of globally interconnected convergent-divergent units may constitute the global workspace (Baars, 1997; Dehaene, Kerzberg, & Changeux, 1998). Such an account would adequately differentiate the global workspace from the various functionally specialized networks, enabling a full dissociation of consciousness and attention (Hsieh, Colas, & Kanwisher, 2011; Webb, Kean, & Graziano, 2016).

## References

Abdelnour, F., Voss, H. U. & Raj, A. (2014). Network diffusion accurately models the relationship between structural and functional brain connectivity networks. Neuroimage, 90, 335–347.

Adesnik, H., Bruns, W., Taniguchi, H., Huang, Z. J., & Scanziani, M. (2012). A neural circuit for spatial summation in visual cortex. *Nature*, *490*(7419), 226–231.

Alexander, D.M., Trengove, C., Sheridan, P., & van Leeuwen, C. (2011). Generalization of learning by synchronous waves: from perceptual organization to invariant organization. Cognitive Neurodynamics, 5, 113-132.

Alexander, D.M., Jurica, P., Trengove, C., …, & van Leeuwen, C. (2013). Traveling waves and trial averaging; the nature of single-trial and averaged brain responses in large-scale cortical signals. Neuroimage, 73, 95–112. doi: 10.1016/j.neuroimage.2013.01.016 .

Arenas, A., Duch, J., Fernández, A., & Gómez, S. (2007). Size reduction of complex networks preserving modularity. New Journal of Physics, 9(6), 176.

Baars, B.J. (1997). In the Theater of Consciousness, The Workspace of the Mind. NY, Oxford University Press.

Bassett, D. S. & Bullmore, E. T. (2017). Small-world brain networks revisited. The Neuroscientist, 23, 499–516.

Bender, E. A., & Williamson, S. G. (2010). *Lists, Decisions and Graphs*. https://cseweb.ucsd.edu/~gill/BWLectSite/Resources/LDGbookCOV.pdf

Blankenship, A.G. & Feller, M.B. (2010). Mechanisms underlying spontaneous patterned activity in developing neural circuits. Nature Reviews Neuroscience, 11(1), 18–29. doi: 10.1038/nrn2759

Bullmore, E. & Sporns, O. (2009). Complex brain networks: graph theoretical analysis of structural and functional systems. Nature Reviews Neuroscience, 10, 186-198.

Butz, M., Wörgötter, F., & van Ooyen, A. (2009). Activity-dependent structural plasticity. Brain Research Reviews, 60(2), 287-305.

Calvo Tapia, C., Makarov, V. A., & van Leeuwen, C. (2020). Basic principles drive self-organization of brain-like connectivity structure. Communications in Nonlinear Science and Numerical Simulation, 82, 105065.

Chapman, A. (2015). Advection on Graphs. In A. Chapman (Ed.), Semi-Autonomous Networks: Effective Control of Networked Systems through Protocols, Design, and Modeling (pp. 3–16). Springer International Publishing.

Cherniak, C. (1994). Component placement optimization in the brain. J Neurosci 1994;14(4):2418–27. doi: 10.1523/JNEUROSCI.14- 04- 02418.

Chklovskii, D. B., Mel, B. W., & Svoboda, K. (2004). Cortical rewiring and information storage. Nature, 431(7010), 782-788.

Corbetta, M., Kincade, J.M., Ollinger, J.M., McAvoy, M.P.,& Shulman, G.L.(2000). Voluntary orienting is dissociated from target detection in human posterior parietal cortex. Nature Neuroscience,3(3),292-297. doi: 10.1038/73009.

Das, A., & Gilbert, C. D. (1999). Topography of contextual modulations mediated by short-range interactions in primary visual cortex. Nature, 399(6737), 655-661.

Dehaene, S., Kerszberg, M., & Changeux, J.P. (1998). A neuronal model of a global workspace in effortful cognitive tasks. Proceedings of the National Academy of Sciences, USA, 95 (24), 14529–14534. doi:10.1073/pnas.95.24.14529.

Doherty, M.J., Tsuji, H, & Phillips, W.A. (2008). The context sensitivity of visual size perception varies across cultures. Perception, 37(9):1426-33. doi: 10.1068/p5946.

Field, D. J., Hayes, A., Hess, R.F, (1993). Contour integration by the human visual system: Evidence for a local "association" field. Vision Research, 33, 173 – 179.

Gong, P. & van Leeuwen, C. (2003). Emergence of scale-free network with chaotic units. Physica A, Statistical mechanics and its applications, 321, 679–688.

Gong, P. & van Leeuwen, C. (2004). Evolution to a small-world network with chaotic units. Europhysics Letters, 67, 328–333.

Gong, P. & van Leeuwen, C. (2009). Distributed dynamical computation in neural circuits with propagating coherent activity patterns. PloS Computational Biology, 5(12), e1000611.

Gorban, A.N., Makarov, V.A. & Tyukin, I.Y. (2019). The unreasonable effectiveness of small neural ensembles in high-dimensional brain. Physics of Life Reviews, 29, 55-88

Hellrigel, S., Jarman, N., & van Leeuwen (2019). Adaptive rewiring of weighted networks. Cognitive Systems Research, 55, 205-218

Hilgetag, C.-C., Burns, G. A., O'Neill, M. A., Scannell, J. W., & Young, M. P. (2000). Anatomical connectivity defines the organization of clusters of cortical areas in the macaque and the cat. Philosophical Transactions of the Royal Society of London B. Biological Sciences, 355, 91–110.

Hopfinger, J.B., Buonocore, M.H., & Mangun, G.R. (2000). The neural mechanisms of top-down attentional control. Nature Neuroscience,3(3),284-291. doi: 10.1038/72999.

Hsieh, P., Colas, J. T., & Kanwisher, N. (2011). Unconscious popout: Attentional capture by unseen feature singletons only when top-down attention is available. Psychological Science, 22, 1220–1226. doi:10.1177/0956797611419302

Hupé, J. M., James, A. C., Payne, B. R., Lomber, S. G., Girard, P., & Bullier, J. (1998). Cortical feedback improves discrimination between figure and background by V1, V2 and V3 neurons. Nature, 394(6695), 784-787.

Jarman, N., Steur, E., Trengove, C., Tyukin, I., & van Leeuwen, C. (2017). Self-organisation of small-world networks by adaptive rewiring in response to graph diffusion. Scientific Reports, 7, 13518.

Jarman, N., Trengove, C., Steur, E., Tyukin, I., & van Leeuwen, C. (2014). Spatially constrained adaptive rewiring in cortical networks creates spatially modular small world architectures. Cognitive Neurodynamics, 8, 479-497.

Jordan, R., & Keller, G. B. (2020). Opposing influence of top-down and bottom-up input on excitatory layer 2/3 neurons in mouse primary visual vortex. Neuron, 108(6), 1194-1206.

Keller, A. J., Dipoppa, M., Roth, M. M., Caudill, M. S., Ingrosso, A., Miller, K. D., & Scanziani, M. (2020). A disinhibitory circuit for contextual modulation in primary visual cortex. Neuron, 108(6), 1181-1193.

Keller, G.B., Bonhoeffer, T., & Hübener, M. (2012) Sensorimotor mismatch signals in primary visual cortex of the behaving mouse. Neuron, 74,(5), 809–815.

Keller, A. J., Dipoppa, M., Roth, M. M., Caudill, M. S., Ingrosso, A., Miller, K. D., & Scanziani, M. (2020). A disinhibitory circuit for contextual modulation in primary visual cortex. *Neuron*, *108*(6), 1181–1193.

Kumar, A., Rotter, S. & Aertsen, A. (2010). Spiking activity propagation in neuronal networks: reconciling different perspectives on neural coding. Nature Reviews Neuroscience, 11, 615–627.

Knott, G. & Holtmaat, A. (2008). Dendritic spine plasticity – current understanding from in vivo studies. Brain Research Reviews 20(58), 282-289. https://doi.org/10.1016/j.brainresrev.2008.01.002.

Krause, M.R. & Pack, C.C. (2014). Contextual modulation and stimulus selectivity in extrastriate cortex. Vision Research, 104, 36-46.

Kwok, H.F., Jurica, P. Raffone, A., & van Leeuwen, C. (2007). Robust emergence of small-world structure in networks of spiking neurons. Cognitive Neurodynamics, 1, 39–51

Latora, V., & Marchiori, M. (2001). Efficient Behavior of Small-World Networks. Physical Review Letters, 87(19), 198701.

Leicht, E. A., & Newman, M. E. (2008). Community structure in directed networks. Physical Review Letters, 100(11), 118703.

Leinweber, M., Ward, D.W., Sobczak, J.M., Attinger, A., & Keller, G.B. (2017). A sensorimotor circuit in mouse cortex for visual flow predictions. Neuron 95, (6), 1420–1432.

Menon, V. & Uddin, L.Q. (2010) Saliency, switching, attention and control: a network model of insula function. Brain Structure and Function, 214(5-6), 655-667. doi: 10.1007/s00429-010-0262-0.

Muller, L., Chavane, F., Reynolds, J., & Sejnowski, T. J. (2018). Cortical travelling waves: Mechanisms and computational principles. *Nature Reviews Neuroscience*, *19*(5), 255–268. https://doi.org/10.1038/nrn.2018.20

Niell, C. M., & Scanziani, M. (2021). How cortical circuits implement cortical computations: Mouse visual cortex as a model. *Annual Review of Neuroscience*, *44*, 517–546.

Oldham, S. & Fornito, A. (2019). The development of brain network hubs. Developmental Cognitive Neuroscience, 36, 100607. doi: 10.1016/j.dcn.2018.12.005.

Papadopoulos L, Kim JZ, Kurths J, Bassett DS. (2017). Development of structural correlations and synchronization from adaptive rewiring in networks of Kuramoto oscillators. Chaos 27(7):073115. doi: 10.1063/1.4 994 819 .

Phillips, W. A, Chapman, K.L.S., & Berry, P.D, (2004). Size perception is less context-sensitive in males. Perception, 33, 79 – 86.

Quian Quiroga, R., Reddy, L., Kreiman, G., Koch., C., & Fried, I. (2005). Invariant visual representation by single neurons in the human brain. Nature, 435(7045), 1102-1107. doi: 10.1038/nature03687.

Raichle, M,E. (2015). The brain's default mode network. Annual Review of Neuroscience, 8(38), 433-47. doi: 10.1146/annurev-neuro-071013-014030.

Raichle, M.E., MacLeod, A.M. Snyder, A.Z., Powers, W.J. Gusnard, D.A. & Shulman, G.L. (2001). A default mode of brain function. Proceedings of the National Academy of Sciences USA, 98(2), 676-682 doi.: 10.1073/pnas.98.2.676

Redgrave, P., Rodriguez, M., Smith, Y., Rodriguez-Oroz, M. C., Lehericy, S., Bergman, H., Agid, Y., DeLong, M. R., & Obeso, J. A. (2010). Goal-directed and habitual control in the basal ganglia: Implications for Parkinson’s disease. *Nature Reviews Neuroscience*, *11*(11), 760–772.

Ren, W., Beard, R. W., & Atkins, E. M. (2007). Information consensus in multivehicle cooperative control. IEEE Control Systems Magazine, 27(2), 71-82.

Rentzeperis, I. & van Leeuwen, C. (2020). Adaptive rewiring evolves brain-like structure in weighted networks. Scientific Reports, 10, Art. No. 6075.

Rentzeperis, I., & van Leeuwen, C. (2021). Adaptive rewiring in weighted networks shows specificity, robustness, and flexibility." Frontiers in Systems Neuroscience, 15, 13.

Rentzeperis, I., Laquitaine, S., & Leeuwen, C. van. (2022). Adaptive rewiring of random neural networks generates convergent–divergent units. *Communications in Nonlinear Science and Numerical Simulation*, *107*, 106135. https://doi.org/10.1016/j.cnsns.2021.106135

Rubinov, M., Sporns, O., van Leeuwen, C., & Breakspear, M. (2009). Symbiotic relationship between brain structure and dynamics. BMC Neuroscience, 10, 55.

Seeley, W.W., Menon, V., Schatzberg, A.F., Keller, J., Glover, G.H., Kenna, H., Reiss, A.L., & Greicius, M.D. (2007). Dissociable intrinsic connectivity networks for salience processing and executive control. Journal of Neuroscience, 27(9), 2349-2356. doi: 10.1523/JNEUROSCI.5587-06.2007.

Shanahan, M. (2012). The brain's connective core and its role in animal cognition, Philosophical Transactions of the Royal Society B, 367(1603), 2704-2714.

Shaw, G. L., Harth, E. & Scheibel, A. B. (1982). Cooperativity in brain function: assemblies of approximately 30 neurons. Experimental Neurology, 77, 324–358.

Sporns, O., and Zwi, J. D. (2004). The small world of the cerebral cortex. Neuroinformatics 2, 145–162.

van den Berg, D., Gong, P., Breakspear, M., & van Leeuwen, C. (2012). Fragmentation: Loss of global coherence or breakdown of modularity in functional brain architecture? Frontiers in Systems Neuroscience, 6, 20. doi: 10.3389/fnsys.2012.00020.

van den Berg, D. & van Leeuwen, C. (2004). Adaptive rewiring in chaotic networks renders small-world connectivity with consistent clusters. Europhysics Letters, 65, 459–464.

Van den Heuvel, M. P., & Sporns, O. (2011). Rich-club organization of the human connectome. Journal of Neuroscience, 31, 15775–15786. doi:10.1523/JNEUROSCI.3539-11.2011.

Völgyi, B., Chheda, S., & Bloomfield, S. A. (2009). Tracer coupling patterns of the ganglion cell subtypes in the mouse retina. *Journal of Comparative Neurology*, *512*(5), 664-687.

Watts, D. J.; Strogatz, S. H. (1998). Collective dynamics of 'small-world' networks. Nature, 393 (6684), 440–442. doi:10.1038/30918.

Webb, T.W., Kean, H.H., & Graziano, M.S.A. (2016). Effects of awareness on the control of attention. Journal of Cognitive Neuroscience, 28, 842–851. doi:10.1162/jocn_a_00931

Yamashita, Y., Fujimura, T., Katahira, K. et al. (2016). Context sensitivity in the detection of changes in facial emotion. Scientific Reports, 6, 27798. https://doi.org/10.1038/srep27798.

Zamora-López, G., Zhou, C., and Kurths, J. (2010). Cortical hubs form a module for multisensory integration on top of the hierarchy of cortical networks. Frontiers in Neuroinformatics, 4, 1.